\documentclass[3p,twocolumn,times]{elsarticle}
\usepackage{ecrc}
\usepackage{multicol}
\volume{00}
\pdfoutput=1
\firstpage{1}
\journalname{Nuclear Physics B Proceedings Supplement}
\runauth{}
\jid{nuphbp}
\jnltitlelogo{Nuclear Physics B Proceedings Supplement}
\usepackage{amssymb}

\usepackage{graphicx}

\usepackage[figuresright]{rotating}

\begin{document}

\begin{frontmatter}



\dochead{}

\title{Determining Reactor Neutrino Flux\tnoteref{label1}}
\author{Jun Cao}
\ead{caoj@ihep.ac.cn}
\address{Institute of High Energy Physics, Beijing 100049, China}

\begin{abstract}
Flux is an important source of uncertainties for a reactor neutrino experiment. It is determined from thermal power measurements, reactor core simulation, and knowledge of neutrino spectra of fuel isotopes. Past reactor neutrino experiments have determined the flux to (2-3)\% precision. Precision measurements of mixing angle $\theta_{13}$ by reactor neutrino experiments in the coming years will use near-far detector configurations. Most uncertainties from reactor will be canceled out. Understanding of the correlation of uncertainties is required for $\theta_{13}$ experiments. Precise determination of reactor neutrino flux will also improve the sensitivity of the non-proliferation monitoring and future reactor experiments. We will discuss the flux calculation and recent progresses.
\end{abstract}


\end{frontmatter}

% \linenumbers

\section{Introduction}
Reactor neutrino experiments have played a critical role in the history of
neutrinos. Among them, Savannah River Experiment~\cite{reines} by Reines and
Cowan in 1956 observed the first neutrino. Chooz~\cite{chooz} determined the
most stringent upper limit of the last unknown neutrino mixing angle
$\sin^22\theta_{13}<0.17$ in 1998. KamLAND~\cite{kamland} observed the first
reactor neutrino disappearance in 2003.

Flux is an important source of uncertainties for a reactor neutrino experiment. It is determined from thermal power measurements, reactor core simulation, and knowledge of neutrino spectra of fuel isotopes. Past reactor neutrino experiments have determined the flux to (2-3)\% precision. In the coming years, three precision experiments on neutrino mixing angle $\theta_{13}$ using reactor neutrinos, Daya Bay~\cite{dyb}, Double Chooz~\cite{dchooz}, and RENO~\cite{reno}, will start operation. All these experiments use near-far detector configurations. Most uncertainties from the reactor will cancel out. The residual error will range from 0.1\% to 0.45\%. However, correlation among reactor cores need better understanding of the error sources. Recently there are increasing interests on non-proliferation monitoring~\cite{nonpro} using ton-level neutrino detectors. A reactor neutrino experiment at an intermediate baseline $\sim$60 km with a giant detector~\cite{dybii} will have rich physics content. Precise determination of neutrino flux will greatly improve the sensitivity of these experiments. Another category of reactor neutrino experiments is $\nu$-electron or $\nu$-nucleus scattering experiments, such as TEXONO~\cite{texono}, MUNU~\cite{munu}, GEMMA~\cite{gemma}, etc. Normally they won't rely on precise neutrino flux.

In this note, we will review the calculation of the reactor neutrino flux and recent progresses on the error analysis.

\section{Calculation of Reactor Neutrino Flux}

\begin{figure}[!htb]
\begin{center}
\includegraphics[width=0.4\textwidth]{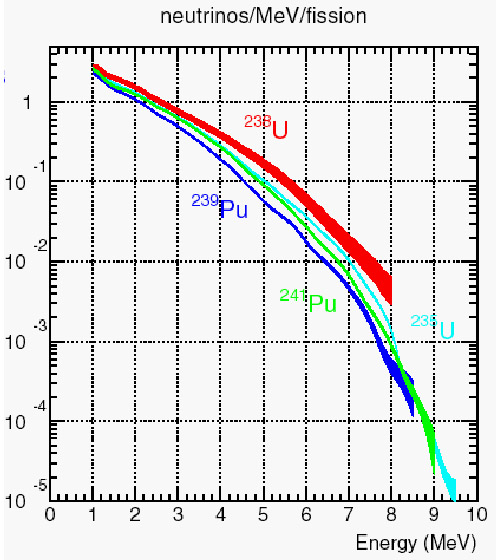}
\caption{Energy spectra of reactor neutrinos.\label{fig:spectra} }
\end{center}
\end{figure}

\begin{figure}[!htb]
\begin{center}
\includegraphics[width=0.4\textwidth]{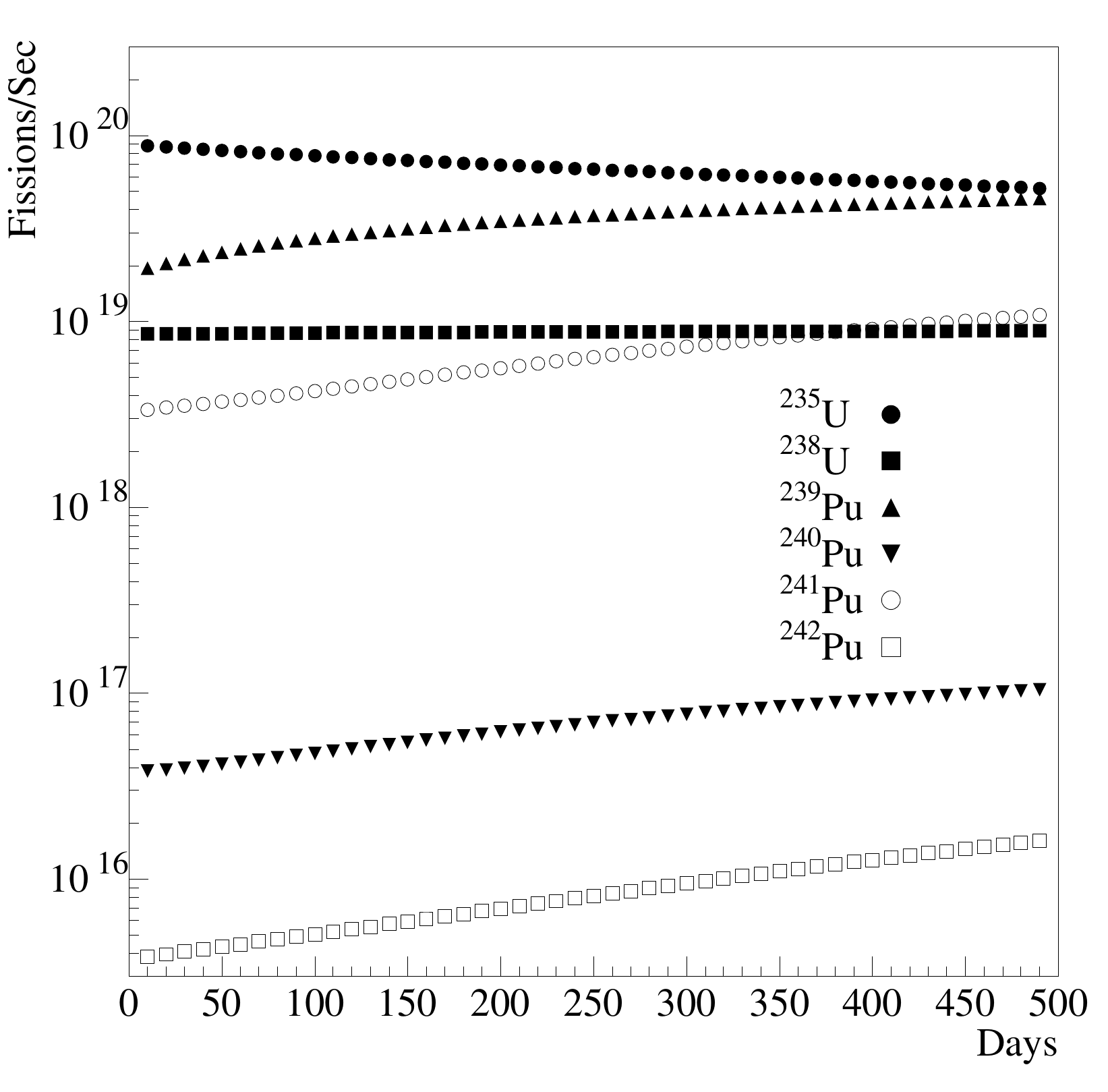}
\caption{Isotope evolution of a typical PWR.\label{fig:evol} }
\end{center}
\end{figure}

Most commercial reactors are Pressurized Water Reactor (PWR) or Boiling Water Reactor (BWR). They are very similar in neutrino flux calculation. We will use PWR as examples in the following. The $^{235}$U enrichment in fresh fuel of a PWR is normally (3-4)\%, and more than 95\% is $^{238}$U. Electron antineutrinos are emitted from  subsequent $\beta$-decays of fission fragments. They are dominated by 4 isotopes,
$^{235}$U, $^{239}$Pu, $^{241}$Pu, and $^{238}$U. Other isotopes contribute
only at 0.1\% level. One can calculate the neutrino energy spectrum of each isotope by summing all fission fragment $\beta$-decay branches. There are a lot of efforts on such studies. However, the fission products are very complex. Due to lack of accurate nuclear data, such calculations carry large uncertainties at the 10\% level.
The most accurate neutrino spectra of the first 3 isotopes were determined at
ILL~\cite{ill} by measuring the $\beta$ spectra of fissioning. The $\beta$ spectra are then converted to neutrino spectra, with an average uncertainty 1.9\%. The $^{238}$U spectrum are calculated theoretically~\cite{vogel238}. They are shown in Fig.~\ref{fig:spectra}.
The isotope concentration in fuel will evolve during reactor operation as $^{235}$U depletes and $^{239}$Pu and $^{241}$Pu breed. The $^{238}$U concentration is relatively stable. Such evolution can be obtained by core simulation. A typical isotope evolution as a function of operation time, in terms of fission rates of the reactor, is shown in Fig.~\ref{fig:evol}.

When we know the fission rates of each isotope from core simulation, and neutrino energy spectrum of each isotope, we can easily get the neutrino flux $S(E_\nu)=\sum_i f_i S_i(E_\nu)$, where $f_i$ is the fission rate of isotope $i$ and $S_i(E_\nu)$ is its neutrino spectrum. However, the fission rates are proportional to the thermal power of the core, which is fluctuating. It is unrealistic to repeat core simulation to reflect the power fluctuation. Normally we scale the neutrino flux to the measured thermal power.
\begin{equation}
S(E_\nu)= \frac{W_{\rm th}}{\sum_i(f_i/F)e_i} \sum_i(f_i/F) S_i(E_\nu) \,,
\end{equation}
where $W_{\rm th}$ is the thermal power, $e_i$ is the energy release per fission for isotope $i$, and $F$ is the sum of $f_i$, thus $f_i/F$ is the fission fraction of each isotope. Among the inputs, the thermal power data is provided by the nuclear power plant. The uncertainty is generally estimated to be (0.6-0.7)\%~\cite{chooz,PaloVerde}. Fission fractions are obtained by core simulation as a function of burn-up. Burn-up is the amount of energy in Mega Watt Days (MWD) released per unit initial mass (ton) of Uranium (TU). The simulated fission fraction carries $\sim$5\% uncertainties from statistics of hundreds of analyses for various codes and various reactors~\cite{chooz,djurcic}. The 5\% fission fraction uncertainties corresponds to $\sim$0.5\% uncertainty in neutrino yield. Energy release per fission varies slightly for different cores at different time due to neutron capture and non-equilibrium products. Average numbers~\cite{james,kopeikin} can be used, with uncertainties of (0.30-0.47)\%. Alternatively we can extract them from the core simulation to accurately reflect the core differences and burn-up effects.

Recently there are studies to include contributions from non-equilibrium isotopes in the core~\cite{kopeikinnon,ruan} as well as that from spent fuel which is temporarily stored adjacent to the core~\cite{sinev,anfp}. These are at sub-percent level and only contribute to the low energy region. Besides fission products, $^{238}$U(n,$\gamma$)$^{239}$U reaction also contributes to the neutrino yield. It is below inverse $\beta$-decay threshold (1.8 MeV) but will contribute significantly to low energy $\nu$-electron scattering experiments~\cite{texono}.

\section{Thermal Power}

The most accurate thermal power measurement is the Secondary Heat Balance method. Detailed description of this measurement can be found, for example, in \cite{kme}. This is an offline measurement, normally done weekly or monthly. The uncertainty is cited as 0.7\% by Chooz and Palo Verde. Primary Heat Balance tests are online thermal power measurement. Normally it is calibrated to the Secondary Heat Balance measurement weekly. Daya Bay power plants control the difference of these two measurements to less than 0.1\% of the full power. These data are good for neutrino flux analysis. To 0.1\% level, it can be taken as the Secondary Heat Balance measurement. The power plants also monitor the ex-core neutron flux, which gives the nuclear power. This monitoring is online, for safety and reactor operation control. It is normally calibrated to the Primary Heat Balance measurement daily. This measurement is less accurate, controlled to be less than 1.5\% of the full power by Daya Bay power plant.

Recently there are a lot of studies on the power uncertainties and instrumentation improvements by the power plants~\cite{djurcic,kme}, with the motivation of power uprates. The power measurements can be more accurate than what were cited in the past reactor neutrino experiments. The uncertainties of the Secondary Heat Balance is dominated by the flow rate measurement. In the past, there are two kinds of widely used flow meters, venturi type flow meters and orifice plate flow meters. Venturi flow meters are used by most US and Japan reactors. The uncertainty is often 1.4\%. It can be as low as 0.7\% if properly calibrated and maintained. But they suffer from fouling effects, which could grow as high as 3\% in a few years. To improve the measurement, ultrasonic flow meters have begun to be in use in some US and Japan reactors. They have uncertainties 0.45\% for Type I and 0.2\% for Type II~\cite{djurcic}. The Orifice plate flow meters are used by French reactors. They have no fouling effects. Typically they have an uncertainty of 0.72\% and could be improved to 0.4\% with laboratory tests. It should be noted that the uncertainties of above flow meters are at the 95\% C.L. (confidence level), as defined in ISO-5167. Unless specified, the thermal power uncertainty given by the power plant is also at 95\% C.L.

An EDF (Electricite de France) N4 reactor with four parallel steam generators, which is the Chooz type, is analyzed in~\cite{kme}. Main components of the uncertainties are shown in Table~\ref{tab:n4}. It is dominated by the discharge coefficient, which is an empirical formula in the flow rate measurement and its uncertainty is specified in ISO 5167-1-2003. The final uncertainty of the thermal power is 0.40\% at 95\% C.L. In this evaluation, it is assumed that the discharge coefficients of the orifice plates in four coolant loops are independent, thus the final uncertainty is statistically reduced. If the discharge coefficients are fully correlated for all four orifice plates, then there is no statistical reduction. The power uncertainty will be 0.37\% at 1$\sigma$ level, which is still significantly smaller than 0.7\% that Chooz used.

\begin{table}[!htb]\small
\caption{Error table for the thermal power measurements for N4 reactor~\cite{kme}.\label{tab:n4} }
\begin{center}
\begin{tabular}{c|c|c}
\hline
& Contribution & Relative fraction \\
Origin of uncertainty &  [MWth] &  [\%] of the  \\
& &  17.2 MWth \\ \hline
Discharge coefficient & 15.33 & 79.57 \\\hline
Differential pressure & 6.33 & 13.57 \\\hline
Steam gen. inlet temp. & 2.81 & 2.68 \\\hline
Primary input & 2.00 & 1.35 \\\hline
Others uncertainties & 2.98 & 3.00 \\\hline\hline
Uncertainty at 95\% C.L. & \multicolumn{2}{|c}{4250$\pm$17.2 MW (0.40\%)} \\\hline
\end{tabular}
\end{center}
\end{table}

\par
The Daya Bay and Ling Ao reactors are all calibrated with the SAPEC system, which is an EDF portable high precision secondary heat balance test system with its own sensors, databases, and data processing, of uncertainty of 0.45\%. The calibration results of Ling Ao reactors can be found in Table~\ref{tab:lakme}~\cite{helishi}. Four tests show differences from 0.031\% to 0.065\%. Such small differences mean that the secondary heat balance system (KME) of the reactor is strongly correlated with the SAPEC system. Actually, the calibrations use the same orifice plates but different pressure transmitters. It proves again that the power uncertainty is dominated by the discharge coefficient. It also shows that Ling Ao KME is in very good agreement with the SAPEC system. The power uncertainty is estimated to be 0.48\% at 95\% C.L. in this comparison. As with the above example, these analyses assumed that the discharge coefficients are uncorrelated, which may not be the case.

\begin{table*}[!htb]
\caption{Comparison of the core power calculation results between KME system and SAPEC system.\label{tab:lakme} }
\begin{center}
\begin{tabular}{c|c|c|c|c|c}
\hline
\multicolumn{2}{c|}{} & Test 1 & Test 2 & Test 3 & Test 4 \\\hline
 & KME(MW) & 2897.1 & 2904.4 & 2908.9 & 2906.9 \\\cline{2-6}
 Thermal & SAPEC(MW) & 2896 & 2903 & 2907 & 2906 \\\cline{2-6}
 Power & Difference(MW) & 1.1 & 1.4 & 1.9 & 0.9 \\\cline{2-6}
  & Difference & 0.038\% & 0.048\% & 0.065\% & 0.031\% \\\hline
  Uncertainty & KME & 0.4806\% & 0.4806\% & 0.4806\% & 0.4806\% \\\cline{2-6}
  Analysis & SAPEC & 0.45\% & 0.45\% & 0.45\% & 0.45\% \\\hline
\end{tabular}
\end{center}
\end{table*}

\section{Core Simulation}

The fission fraction of fuel isotopes are obtained by core simulation. Qualified core simulation codes are normally licensed, and not available to scientific collaborations. The core simulation also needs a lot of information from the power plant as inputs. Fortunately, the fission fraction can be extracted as a by-product of the refueling calculation required by the power plant, as a function of burn-up. The uncertainties of the obtained fission fraction depends on the simulation code. It only slightly depends on the inputs such as temperature, pressure, Boron concentration, etc.~\cite{miller}, as tested by the simulation code ROCS. The uncertainties of the simulation can be studied by comparing the measured and calculated concentration of fuel isotopes sampled at different burn-up when refueling. These studies are normally a part of the qualification of the licensed simulation code. In ref.~\cite{djurcic}, 159 such studies for various codes and various reactors in US and Japan haven been collected and analyzed. On average, the simulated concentration of isotopes have uncertainties of $\sim$4\% for $^{235}$U, $\sim$5\% for $^{239}$Pu, $\sim$6\% for $^{241}$Pu, and $\sim$0.1\% for $^{238}$U. Assuming the simulated neutron flux in the core is not affected by small variations of isotope concentration, the fission rate is proportional to the isotope concentration. Due to the strong constraint of the total thermal power, simulated concentrations of the foure isotopes are not independent. A 5\% error on the isotope concentration corresponds to a $\sim$0.5\% uncertainty on the detected neutrino rate via inverse $\beta$-decay reaction. There also are large core-to-core correlations for reactors simulated with the same code.

The energy release per fission $e_i$ in Eq.~1 is defined as the energy absorbed in the reactor per fission event. Neutrinos will take away energy from the total energy released in a nuclear fission. Some fission fragments of long lifetime will not reach equilibrium in a short time, thus part of the energy is not released. Each fission will produce 2-3 neutrons while only one neutron will be used to maintain the chain reaction at stable state in reactor running. Other neutrons are absorbed and release energy via neutron capture. As the fuel composition evolves, the contribution of neutron capture will change. Taking the above three corrections into account, $e_i$ varies slightly for different reactors and it changes with time. Average $e_i$'s were evaluated for typical reactor configurations in refs.~\cite{james} and \cite{kopeikin}. A more accurate estimation of $e_i$  as a function of burn-up can be obtained from core simulation, where these corrections are included automatically.
\begin{table}[!htb]
\caption{Energy release per fission in MeV from refs.~\cite{james} and \cite{kopeikin}.\label{tab:erpf} }
\begin{center}
\begin{tabular}{ccc}
\hline\hline
Isotopes & James & Kopeikin \\ \hline
$^{235}$U & 201.7$\pm$0.6 & 201.92$\pm$0.46 \\
$^{238}$U & 205.0$\pm$0.9 & 205.52$\pm$0.96 \\
$^{239}$Pu & 210.0$\pm$0.9 & 209.99$\pm$0.60 \\
$^{241}$Pu & 212.4$\pm$1.0 & 213.60$\pm$0.65 \\\hline
\end{tabular}
\end{center}
\end{table}

\section{Neutrino Spectra}
Electron antineutrinos are emitted from the subsequent $\beta$-decays of fission fragments. Due to the lack of data for the $\beta$-decays of the complex fission products, theoretical calculations of the neutrino spectra of isotopes carry large uncertainties. ILL~\cite{ill} measured the $\beta$ spectra of fissioning of $^{235}$U, $^{239}$Pu, and $^{241}$Pu by thermal neutrons, and converted them to neutrino spectra. The normalization error is estimated to be 1.9\%. Spectrum shape error is from 1.34\% at 3 MeV to 9.2\% at 8 MeV, as shown in Fig.~\ref{fig:spectra}. $^{238}$U can not fission with thermal neutrons. Its spectrum relies on theoretical calculation. The uncertainty is estimated to be 10\%~\cite{vogel238}. Normally $^{238}$U contributes (7-10)\% of fissions in a PWR. The calculated neutrino counting rate and spectra were verified by Bugey and
Bugey-3~\cite{bugey}. The normalization error is further lowered to 1.6\%, which was used by Chooz.

\section{Non-equilibrium Isotopes and Spent Fuel}
\begin{figure}[!htb]
\begin{center}
\includegraphics[width=0.4\textwidth]{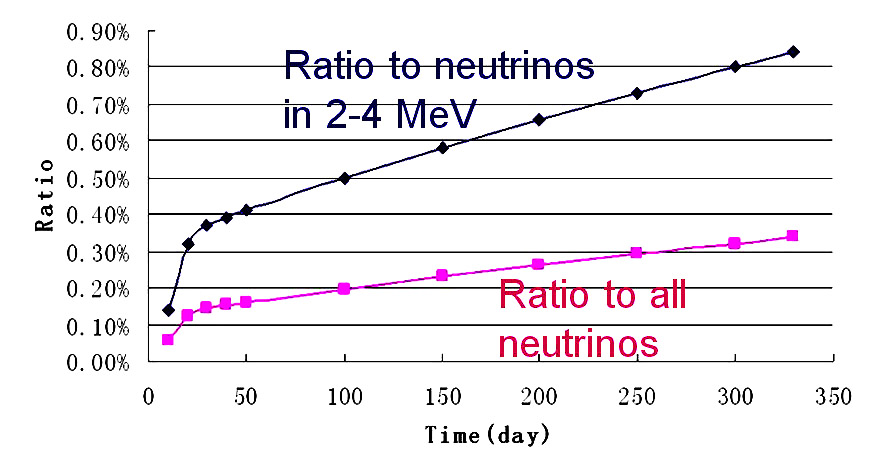}
\caption{Contribution of non-equilibrium isotopes as a fraction of the total neutrinos, weighted by the inverse $\beta$-decay cross section~\cite{ruan}.\label{fig:noneq} }
\end{center}
\end{figure}

The ILL spectra are derived after 1.5 days exposure time with thermal neutron. Thus, long-lived fission fragments have not reached equilibrium. In a real reactor, these fission products will accumulate and contribute to the neutrino flux. Chooz estimated this contribution to be $\sim$0.3\% on the average and ignored it in the detailed analysis due to its small size comparing to other errors. Six chains have been identified in \cite{kopeikinnon}, with half lives from 10 hours to 28 years. They only contribute to the low energy region. Further studies show that for a typical PWR, on average these contributions are $\sim$0.2\% of total neutrino detection rate via inverse $\beta$-decay~\cite{ruan}. In the 2-4 MeV region, it increases to 0.8\% after one year's accumulation, as shown in Fig.~\ref{fig:noneq}

Spent fuel is normally stored temporarily adjacent to the core. The storage could be as long as 10 years. Similar to the non-equilibrium contributions, the long-lived fission fragments in the spent fuel will contribute to the neutrino flux. A PWR is normally refueled very 12-18 months. The spent fuel from one refueling will contribute $\sim$0.2\% of the total neutrino rate after first several days.

\section{Conclusion}
Before the 1980's, the reactor neutrino flux was determined to an uncertainty of 10\%. With a lot of efforts, especially by ILL, Bugey, Chooz, Palo Verde, etc., the uncertainty has improved to (2-3)\%. Motivated by high precision neutrino measurements by scientific collaborations and power uprates by the power plants, we have more accurate thermal power. The uncertainty could be lowered from 0.7\% to 0.4\%. Small corrections from non-equilibrium isotopes and spent fuel, as well as energy release per fission are studied in detail. We also have a global picture of uncertainties of fission rate simulations. However, there is no new data on neutrino spectra of fuel isotopes. For single-detector experiments, the neutrino spectrum uncertainty of about 2\% will dominate. The next $\theta_{13}$ experiments with near-far relative measurements will suffer little from reactor flux uncertainties, which is estimated to be 0.1\% to 0.45\%, depending on the experiment layout. The correlation among reactor uncertainties is important for $\theta_{13}$ experiments since correlated errors will cancel out. Meanwhile, high precision detector and high statistics at the near detector of these experiments may help to improve the knowledge of neutrino spectra.

\par
Since the time that this manuscript was written, it has been suggested~\cite{anomaly} that the antineutrino spectra for all of the relevant fission isotopes ($^{235}$U, $^{238}$U, $^{239}$Pu, and $^{241}$Pu) are several percent larger than those of ref.~\cite{ill,vogel238}. This issue requires further attention.

\bibliographystyle{elsarticle-num}
\bibliography{}

\begin{thebibliography}{50}

\bibitem{reines} C. L. Cowan {\it et al.} Science 124, 103 (1956); F. Reines and
C. L. Cowan, Jr., Nature 178, 446 (1956).

\bibitem{chooz} M. Apollonio {\it et al.}, Eur. Phys. J. C27, 331 (2003)

\bibitem{kamland}  T. Araki {\it et al.}, Phys. Rev. Lett. {\bf 94}, 081801 (2005).

\bibitem{dyb} Daya Bay Proposal, hep-ex/0701029

\bibitem{dchooz} Double Chooz Proposal, hep-ex/0606025

\bibitem{reno} RENO proposal, arxiv:1003.1391

\bibitem{nonpro} For examples, A. Bernstein {\it et al.}, J. Appl. Phys. {\bf 91}, 4672 (2002); J. Appl. Phys. {\bf 103}, 074905 (2008).

\bibitem{dybii} L. Zhan {\it et al.}, Phys.\ Rev.\ D{\bf 78}, 111103 (2008); Phys.\ Rev.\ D{\bf 79}, 073007 (2009);

\bibitem{texono} H.B.~Li {\it et al.}, Phys. Rev. Lett. 90, 131802 (2003); H.T Wong {\it et al.}, Phys. Rev. {\bf D}75, 012001 (2007).

\bibitem{munu} Z. Daraktchieva {\it et al.} Phys. Lett. B564, 190 (2003).

\bibitem{gemma} A.G. Beda {\it et al.}, Phys. Atom. Nucl. {\bf 70}, 1873 (2007).

\bibitem{ill} K. Schreckenbach, G. Colvin and F. von Feilitzsch,
Phys. Lett. B160, 325 (1985); F. von Feilitzsch and K. Schreckenbach, Phys. Lett. B118 (1982); A. A. Hahn {\it et al.}, Phys. Lett. B218, 365 (1989).

\bibitem{vogel238} P. Vogel {\it et al.}, Phys. Rev. C{\bf 24}, 1543 (1981).

\bibitem{PaloVerde}
F.~Boehm {\it et al.}, Phys.\ Rev.\ D{\bf 62}, 072002 (2000).

\bibitem{djurcic} Z. Djurcic {\it et al.}, J. Phys. G: Nucl. Part. Phys. {\bf 36}, 045002 (2009).

\bibitem{james} M. F. James, J. Nucl. Energy 23, 517 (1969).

\bibitem{kopeikin} V. Kopeikin {\it et al.}, Phys. Atom. Nucl., 67, 1892 (2004).

\bibitem{kopeikinnon} V. Kopeikin {\it et al.}, Phys. Atom. Nucl. {\bf 64}, 849 (2001).

\bibitem{ruan} X.C. Ruan {\it et al.}, private communicatioin.

\bibitem{sinev} V. Kopeikin {\it et al.}, Phys. Atom. Nucl. {\bf 69}, 185 (2006).

\bibitem{anfp} Feng-Peng An {\it et al.}, Chinese Physics C{\bf 33}, 711 (2009).

\bibitem{kme} {\it Imrpoving Pressurized Water Reactor Performance Through Instrumentation: Application Case of Reducing Uncertainties on Thermal Power}, EPRI report prepared by Electricite de France, 2001; {\it Application of Orifice Plates for Measurement of Feedwater Flow}, EPRI report prepared by Electricite de France, 2001.

\bibitem{helishi} C. Xu {\it et al.}, Chinese Journal of Nuclear Science and Engineering {\bf 23}, 26 (2003).

\bibitem{miller} L. Miller, Ph.D thesis, 2001, Stanford University.

\bibitem{bugey} Y.~Declais {\it et al.}, Phys.\ Lett.\ {\bf B338}, 383 (1994).
B.~Ackar {\it et al.}, Nucl.\ Phys.\ {\bf B434}, 503 (1995); B. Ackar {\it et
al.}, Phys.\ Lett.\ {\bf B374}, 243 (1996).

\bibitem{anomaly} Th. A. Mueller, {\it et al.}, Phys. Rev. {\bf C}83, 054615 (2011).

\end{thebibliography}



\end{document}